\documentclass[aps,prl,twocolumn,amsmath,amssymb,nofootinbib,superscriptaddress]{revtex4-1}

\usepackage{graphicx}
\usepackage{mathrsfs}
\usepackage[usenames,dvipsnames]{color}
\usepackage{hyperref}
\hypersetup{colorlinks = true, linkcolor = [rgb]{0.19411,0.51882,0.667058}, urlcolor = [rgb]{0.125490,0.29542,0.1647058}, citecolor = [rgb]{0.75882,0.37411,0.14117}}
\usepackage{comment}
\usepackage{mathtools}
\usepackage{notes2bib}
\usepackage[normalem]{ulem}
\usepackage{soul}
\setstcolor{red}
\usepackage{cleveref}

\begin{document}

\bibliographystyle{apsrev}

\title{Enhanced signal-to-noise ratio in \break Hanbury Brown Twiss interferometry by parametric amplification}

\author{Xiaoping Ma}
\email[]{cherry901115@hotmail.com}
\affiliation{Department of Physics, Ocean University of China, Qingdao, 266100, China}
\affiliation{Hearne Institute for Theoretical Physics and Department of Physics \& Astronomy, Louisiana State University, Baton Rouge, LA 70803, United States}

\author{Chenglong You}
\affiliation{Hearne Institute for Theoretical Physics and Department of Physics \& Astronomy, Louisiana State University, Baton Rouge, LA 70803, United States}

\author{Sushovit Adhikari}
\affiliation{Hearne Institute for Theoretical Physics and Department of Physics \& Astronomy, Louisiana State University, Baton Rouge, LA 70803, United States}

\author{Yongjian Gu}
\affiliation{Department of Physics, Ocean University of China, Qingdao, 266100, China}

\author{Jonathan P. Dowling}
\affiliation{Hearne Institute for Theoretical Physics and Department of Physics \& Astronomy, Louisiana State University, Baton Rouge, LA 70803, United States}
\affiliation{NYU-ECNU Institute of Physics at NYU Shanghai,
3663 Zhongshan Road North, Shanghai, 200062, China}

\author{Hwang Lee}
\affiliation{Hearne Institute for Theoretical Physics and Department of Physics \& Astronomy, Louisiana State University, Baton Rouge, LA 70803, United States}

\date{\today}

\begin{abstract}
The Hanbury Brown Twiss (HBT) interferometer was proposed to observe intensity correlations of starlight to measure a star's angular diameter. As the intensity of light that reaches the detector from a star is very weak, one cannot usually get a workable signal-to-noise ratio. We propose an improved HBT interferometric scheme introducing optical parametric amplifiers into the system, to amplify the correlation signal, which is used to calculate the angular diameter. With the use of optical parametric amplifiers, the signal-to-noise ratio can be increased up to 400 percent.

\end{abstract}

\maketitle

 In 1956, a new type of optical interferometer was proposed by Hanbury Brown and Twiss to measure correlations in intensity fluctuations\ \cite{Twiss1956,Brown1956}. They demonstrated the experiment with two photomultipliers placed in the far-field zone of a radiation source to measure the intensity of the collected light and the correlation between the two intensity signals from the two photomultipliers. 
 
 
The Hanbury Brown Twiss (HBT) interferometer was initially developed to measure the angular size of stars. But, for over half a century, the HBT interferometer has also played a major role in understanding the dual wave-particle nature of light by motivating both theoretical and experimental research on the coherence properties of optical fields\ \cite{R1963,Glauber1963,Mandel1965,Perina1985,Baym1993,Singer2013,Loaiza2016,Bai2017}. Moreover, besides the study of optical fields, the HBT interferometer has also been applied in other physical systems. Baym discussed the basic physics of intensity interferometry and its application in high energy nuclear physics, condensed matter physics and atomic physics\ \cite{Baym1998}. Jeltes, \emph{et al}. studied differences between fermionic and bosonic HBT interferometers \cite{Jeltes2007}. Recently, Campagnan, \emph{et al}. introduced a HBT interferometer realized with anyons, which can directly probe entanglement and fractional statistics of initially uncorrelated particles \cite{campagnano2012}. Other interesting applications can be found in Refs.~\cite{silva2016,schellekens2005,hassinen2011,silva2017}.

The first experiment with a HBT interferometer was done on the star Sirius to measure its angular diameter. This star was particularly chosen because it was the only star bright enough to give a workable signal-to-noise ratio (SNR)\ \cite{Twiss1956}. Now the question arises, is it possible to apply HBT interferometry for stars that are much less brighter than Sirius? We answer this question in the affirmative by proposing an interferometric scheme that consists of optical parametric amplifiers (OPAs) applied to the HBT interferometer to obtain an amplified SNR. To be more specific, the original HBT interferometer collects light produced by independent sources on the disc of a star and the intensity is detected at two different locations on Earth, which are multiplied and integrated into a correlator to get the correlation function and finally the angle information. Our idea is to amplify the signal (starlight) before it is passed into the correlator by using two OPAs. The OPAs boosts the input photon number increasing the intensity of the measured star light and ultimately the correlation signal amplitude of the HBT interferometer. The net effect is the increase in SNR. Although the noise also increases, the increase in noise is lower than the signal. We show that if the starlight is very weak (or the mean photon number is very low) the SNR can even increase up to 400 percent. Thus, the new HBT interferometric scheme is helpful for measuring a star whose intensity at the detector is low. 

\emph{The HBT interferometer}.---The HBT interferometer is an improvement over the Michelson stellar interferometer. The Michelson stellar interferometer measures the correlation of the electric field in order to measure the angular diameter of the stars, whereas, the HBT interferometer measures the intensity correlation of the light to measure the angular diameter.

\begin{figure}[h]
\centering
\includegraphics[width=0.3\textwidth]{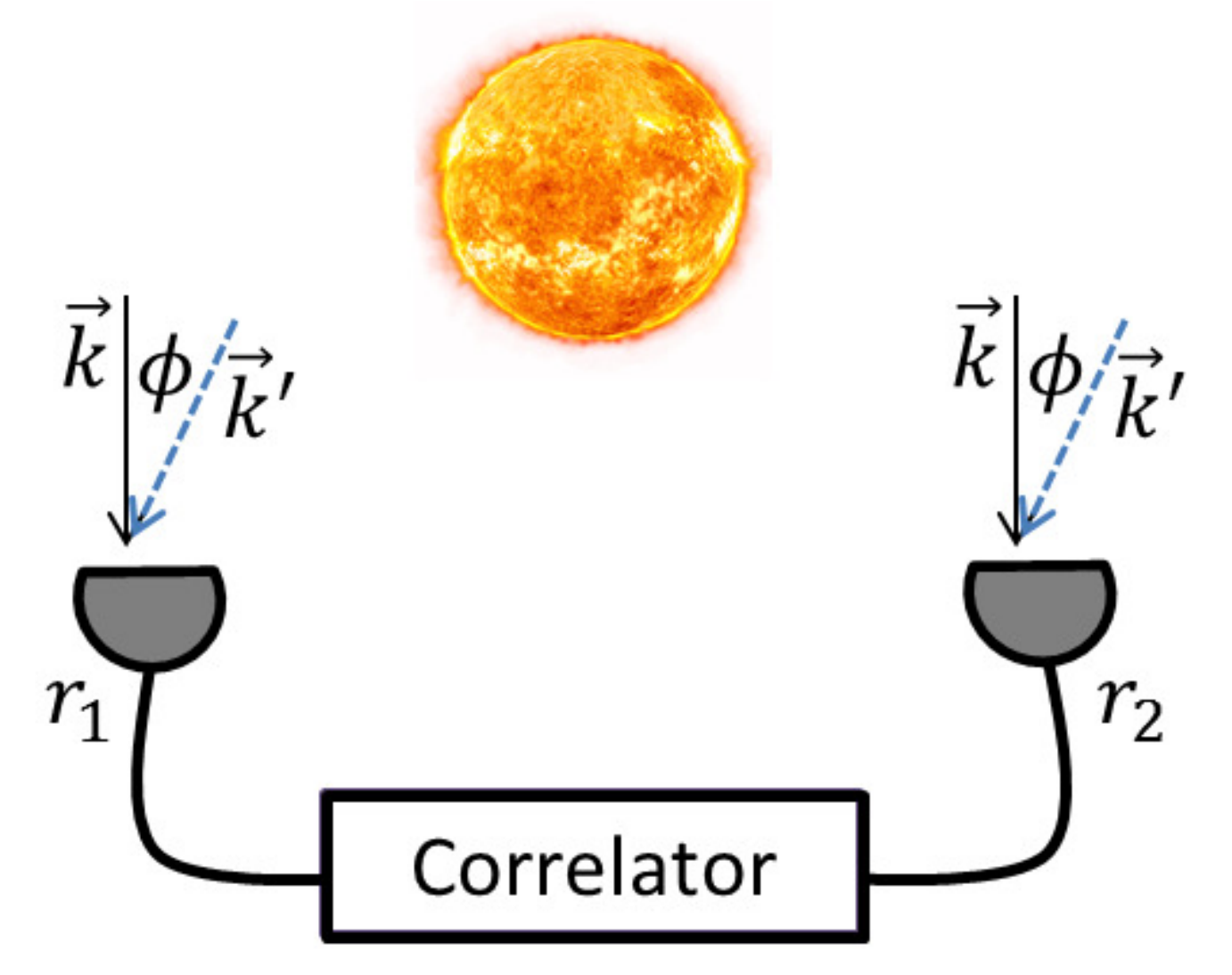}
\caption{\label{fig1}  Schematic diagram of a HBT interferometer. $\vec{k}$ and $\vec{k}^{'}$ are the wave vectors of the two rays. The intensities are measured at ${r_1}$ and ${r_2}$ by two detectors and the signals are combined in an electronic correlator, which calculates the second order correlation function. }
\end{figure}

The basic idea behind the HBT interferometer is shown in Fig.~\ref{fig1}. Let $\vec k$ and $\vec k^{'}$ be the wave vectors of two light beams produced by independent sources on the disc of a star and $\phi$ be the angle between the emitted light. Assuming, sources $\vec k$ and $\vec k^{'}$ produces electric fields ${E_k}{e^{i\vec k \cdot \vec r}}$ and ${E_{{k^{'}}}}{e^{i\vec k^{'} \cdot \vec r}}$, the total amplitudes at ${r_1}$ and ${r_2}$ can be written as
\begin{equation}\label{eq1}
\begin{array}{l}
E\left( {{{\vec r}_1}} \right) = {E_k}{e^{i\vec k \cdot {{\vec r}_1}}} + {E_{{k^{'}}}}{e^{i{{\vec k}^{'}} \cdot {{\vec r}_1}}}\\
E\left( {{{\vec r}_2}} \right) = {E_k}{e^{i\vec k \cdot {{\vec r}_2}}} + {E_{{k^{'}}}}{e^{i{{\vec k}^{'}} \cdot {{\vec r}_2}}}.
\end{array}
\end{equation}
The intensities at ${r_1}$ and ${r_2}$ can be measured with two detectors. The signals which are proportional to the intensities are multiplied and integrated in a correlator. And finally, we obtain the correlation function from the correlator as
\begin{equation}\label{eq2}
\begin{split}
&C_{0} = \langle {E^ * }( {\vec r_1} ){E^ * }( {\vec r_2} )E( {\vec r_1} )E( {\vec r_2} ) \rangle \\&
{\rm{ = }}\langle {{{( {{{\left| {{E_k}} \right|}^2} + {{| {E_{k^{'}}} |}^2}})}^2}} \rangle  + 2\langle {{{| {{E_k}} |}^2}} \rangle \langle {{{| {E_{k^{'}}} |}^2}} \rangle  \cos ( {\vec k - {{\vec k}^{'}}} ) \cdot ( {{{\vec r}_1} - {{\vec r}_2}})
\\&{\rm{ = }}\langle {{{( {{{\left| {{E_k}} \right|}^2} + {{| {E_{k^{'}}} |}^2}})}^2}} \rangle  + 2\langle {{{| {{E_k}} |}^2}} \rangle \langle {{{| {E_{k^{'}}} |}^2}} \rangle  \cos ( k{r_0}\phi), 
\end{split}
\end{equation}
where, $k = | {\vec k} | = | {{{\vec k}^{'}}} |$ and ${{r}_0} = |{{\vec r}_1} - {{\vec r}_2}|$ is the magnitude of the vectorial distance between the two detectors. By varying the separation of the detectors, we can learn the angle between the two rays from Eq.~(\ref{eq2}), and from that, we can get the physical size of the star. As the intensity of each light ray is proportional to the photon number, the correlation function in Eq.~(\ref{eq2}) can also be written as 
\begin{equation}\label{eq3}
{C_0} = \left\langle {{{\hat n}_k}^2} \right\rangle  + \left\langle {{{\hat n}_{{k^{'}}}}^2} \right\rangle  + 2\left\langle {{{\hat n}_k}} \right\rangle \left\langle {{{\hat n}_{{k^{'}}}}} \right\rangle \left( {1 + \cos \left( {k{r_0}\phi } \right)} \right).
\end{equation}
Only the last term in the above equation contains the angle information, and hence we can simplify the correlation function to eliminate the impact of the background as 
\begin{equation}\label{eq4}
C_0^{'} = {C_0} - {C_{0,\rm{DC}}}
\end{equation}
where, ${C_{0,\rm{DC}}}$ is the constant term in the original correlation function. 

The noise is defined as
$\Delta C = \sqrt {\langle {{{\hat C}^2}}\rangle  - {\langle {\hat C} \rangle ^2}}$.
In the HBT interferometer, ${{\hat C}_0} = \hat I\left( {{{\vec r}_1}} \right)\hat I\left( {{{\vec r}_2}} \right)$, and the noise is given as
\begin{equation}\label{eq5}
\begin{split}
&(\Delta {C_0})^2 = \langle {{{\hat n}_k}^4} \rangle  - \langle {{{\hat n}_k}^2} \rangle ^2 + 8\langle {{{\hat n}_k}^3} \rangle \langle {{{\hat n}_{{k^{'}}}}} \rangle  - 4\langle {{{\hat n}_k}^2}\rangle \langle {{{\hat n}_k}} \rangle \langle {{{\hat n}_{{k^{'}}}}} \rangle \\&
 + 8\langle {{{\hat n}_k}} \rangle \langle {{{\hat n}_{{k^{'}}}}^3} \rangle  - 4\langle {{{\hat n}_k}} \rangle \langle {{{\hat n}_{{k^{'}}}}} \rangle \langle {{{\hat n}_{{k^{'}}}}^2} \rangle  + 16\langle {{{\hat n}_k}^2} \rangle \langle {{{\hat n}_{{k^{'}}}}^2} \rangle+ \langle {{{\hat n}_{{k^{'}}}}^4} \rangle  \\& 
 - {\langle {{{\hat n}_{{k^{'}}}}^2} \rangle ^2}- 6{\langle {{{\hat n}_k}} \rangle ^2}{\langle {{{\hat n}_{{k^{'}}}}} \rangle ^2} 
 +4[2(\langle {{{\hat n}_k}^3} \rangle \langle {{{\hat n}_{{k^{'}}}}} \rangle + 2\langle {{{\hat n}_k}^2} \rangle \langle {{{\hat n}_{{k^{'}}}}^2} \rangle\\&
 +\langle {{{\hat n}_k}} \rangle \langle {{{\hat n}_{{k^{'}}}}^3} \rangle)-\langle {{{\hat n}_k}^2}\rangle \langle {{{\hat n}_k}} \rangle \langle {{{\hat n}_{{k^{'}}}}} \rangle-2\langle {{{\hat n}_k}} \rangle ^2{\langle {{{\hat n}_{{k^{'}}}}} \rangle ^2}-\langle {{{\hat n}_k}} \rangle\\&  \times \langle {{{\hat n}_{{k^{'}}}}} \rangle\langle {{{\hat n}_{{k^{'}}}}^2} \rangle]\cos ( {k{r_0}\phi } )+2(\langle {{{\hat n}_k}^2} \rangle \langle {{{\hat n}_{{k^{'}}}}^2} \rangle - 2{\langle {{{\hat n}_k}} \rangle ^2}{\langle {{{\hat n}_{{k^{'}}}}} \rangle ^2})\\&\times\cos ( {2k{r_0}\phi } ).
\end{split}
\end{equation}
In order to remove the influence of the angle, we define a new noise as
\begin{equation}\label{eq6}
\Delta C_0^{'} = \Delta {C_{0,\rm{avg}}}
\end{equation}
where, $\Delta {C_{0,\rm{avg}}}$ is the average of the original noise, meaning all the terms including $\phi$ in the above equation are removed.\\ 
Note that for the thermal state
\begin{equation}\label{eq7}
\hat \rho  = \frac{1}{{1 + {n_\text{th}}}}\sum\limits_{n = 0}^\infty  {{{\left( {\frac{{{n_\text{th}}}}{{1 + {n_\text{th}}}}} \right)}^n}} \left| n \right\rangle \langle n|,
\end{equation}
we have,
\begin{equation}\label{eq8}
\begin{split}
&\left\langle {{{\hat n}^2}} \right\rangle  = 2{{n_\text{th}}^2} + {n_\text{th}},\\&
\left\langle {{{\hat n}^3}} \right\rangle  = {n_\text{th}}\left( {1 + 6{n_\text{th}} + {{n_\text{th}}^2}} \right),\\&
\left\langle {{{\hat n}^4}} \right\rangle  = {n_\text{th}}\left( {1 + 14{n_\text{th}}+ 36{{n_\text{th}}^2} + 24{{n_\text{th}}^3}} \right),
\end{split}
\end{equation}
where, ${n_\text{th}}$ is the mean photon number of the thermal state. 

For the sake of simplicity, we represent the mean photon number of light rays with wave vectors, $\vec {k}$ and $\vec k^{'}$ as $\bar{n}$ and $\bar{m}$ respectively. Using this, the correlation function of the HBT interferometer is given by
\begin{equation}\label{eq9}
{C_0^{'}} = 2\bar{n}\bar{m} \cos ( {k{r_0}\phi } ),
\end{equation}
and, the noise is given by 

\begin{equation}\label{eq10}
\begin{split}
&(\Delta {C_0^{'}})^2 = \bar{n} + 13{\bar{n}^2} + 32{\bar{n}^3} + 20{\bar{n}^4} + \bar{m}+ 13{\bar{m}^2} \\& + 32{\bar{m}^3} + 20{\bar{m}^4}+ 32\bar{n}\bar{m} + 76{\bar{n}^2}\bar{m}+ 76\bar{n}{\bar{m}^2} \\&
+ 40{\bar{n}^3}\bar{m}+ 40\bar{n}{\bar{m}^3} +70{\bar{n}^2}{\bar{m}^2}.
\end{split}
 \end{equation}

\begin{figure}[h]
\centering
\includegraphics[width=0.25\textwidth]{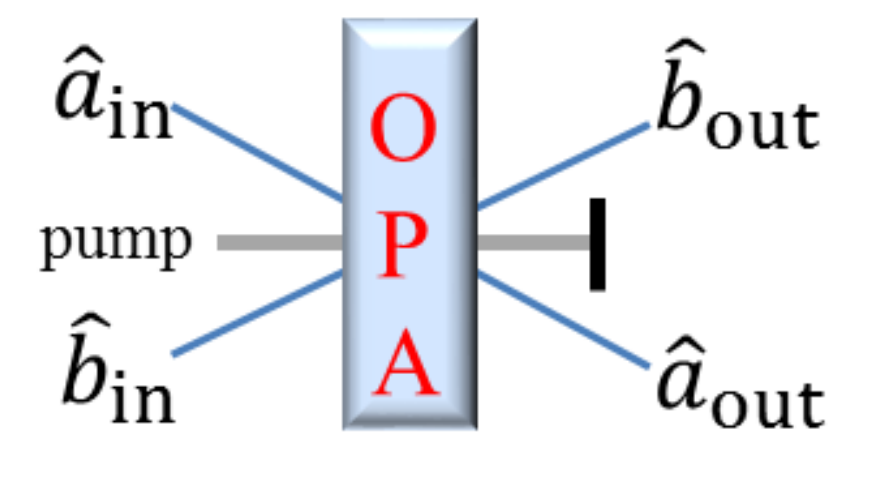}
\caption{\label{fig2}Schematic diagram of an ideal OPA. It consists of a pump beam and two input modes denoted by annihilation operators, $\hat a_\text{in}$ and $\hat b_\text{in}$, incident on the OPA. The $\hat a_\text{out}$ and $\hat b_\text{out}$ are the two output modes. }
\end{figure}

\emph{Model of an optical parametric amplifier}.---An optical parametric amplifier, abbreviated as OPA, is a light source that emits light of variable wavelengths by an optical parametric amplification process. As depicted in Fig.\ \ref{fig2}, an OPA is a device with two input modes, $\hat a_\text{in}$ and $\hat b_\text{in}$ and two output modes, $\hat a_\text{out}$ and $\hat b_\text{out}$, which performs the mode evolution as shown below \cite{Yurke1986}

\begin{equation}\label{eq11}
{\hat T_{\text{OPA}}} = \left( {\begin{array}{*{20}{c}}
{{u}}&{{v}}\\
{v^*}&{{u}}
\end{array}} \right),
\end{equation}
with, ${u} = \cosh{g} $, ${v} = {e^{i{\theta}}}\sinh{g}$, and ${\theta}$ and ${g}$ is the phase shift and parametrical strength in the OPA. 

After propagation through the OPA, the relation between input and output modes are given by
\begin{equation}\label{eq12}
\left( {\begin{array}{*{20}{c}}
{{{\hat a}_\text{out}}}\\
{\hat b_\text{out}^\dag }
\end{array}} \right) = \hat T_{\text{OPA}}\left( {\begin{array}{*{20}{c}}
{{{\hat a_\text{in}}}}\\
{\hat b_\text{in}^\dag }
\end{array}}
\right).
\end{equation}

We can always choose the parameters in the propagation process in Eq.~(\ref{eq12}). Without loss of generality, assuming, ${\theta} = 0$, we can write the relation between the input and output modes as 
\begin{equation}\label{eq13}
\begin{array}{*{20}{c}}
\hat{a}_\textrm{out} = {\hat a_\textrm{in}} \cosh{g} + \hat{b}_\text{in}^\dag \sinh{g}
\\
\hat{b}_\textrm{out}^\dag  = \hat{a}_\textrm{in} \sinh{g}  + \hat b_\textrm{in}^\dag \cosh{g}. 
\end{array}
\end{equation}

\begin{figure}[h]
\centering
\includegraphics[width=0.35\textwidth]{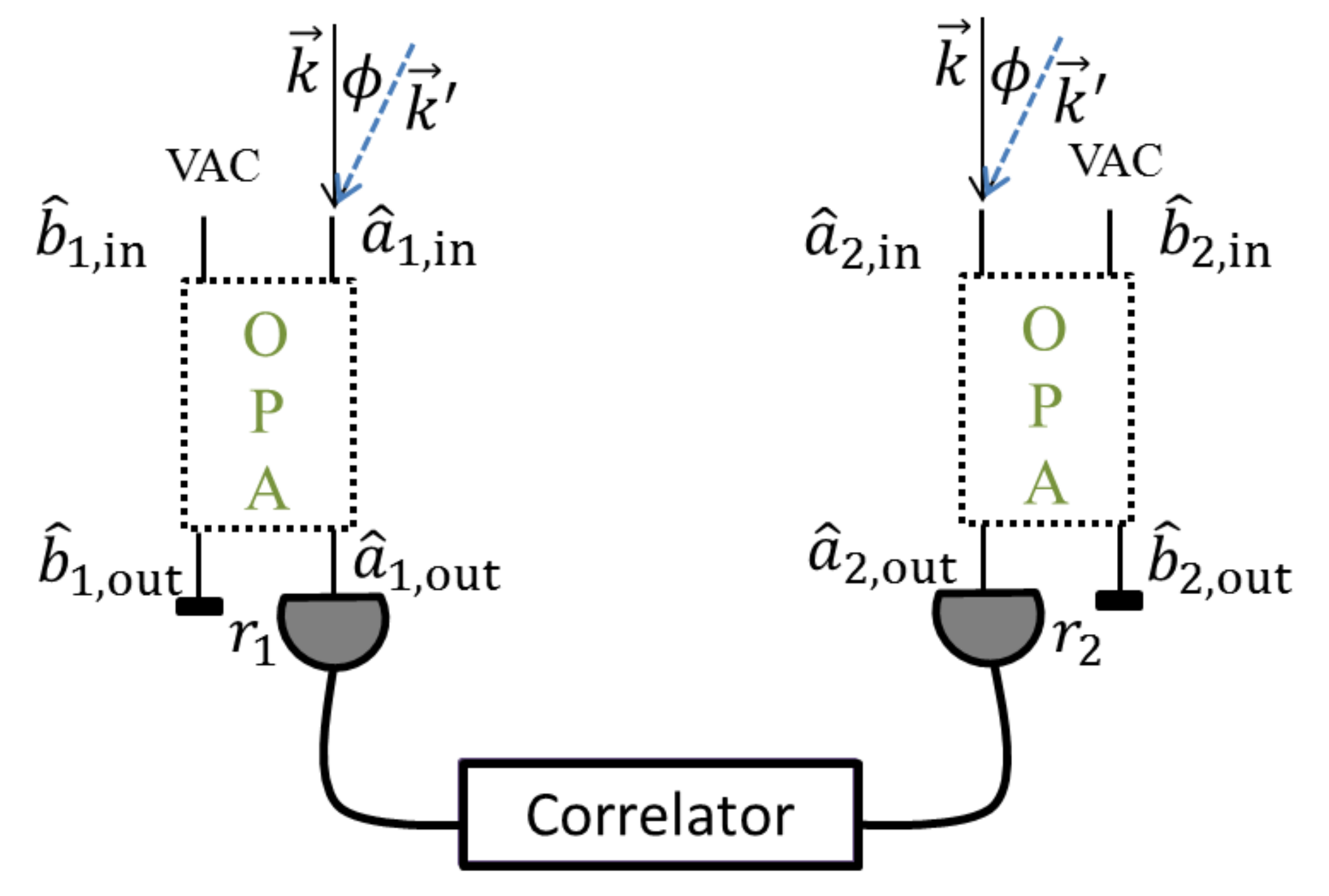}
\caption{\label{fig3} Schematic diagram of our improved HBT interferometer. There are two OPAs placed at ${r_1}$ and ${r_2}$. One input of the OPAs is the star light and the other input is the vacuum state. Two outputs given by, ${\hat a_\text{1,out}}$ and ${\hat a_\text{2,out}}$, are detected and the signals are injected into the correlator.}
\end{figure}

If we inject a thermal state to the upper mode and a vacuum state to the lower mode, then after the propagation, we have
 \begin{equation}\label{eq14}
 \langle {{{\hat n}_\textrm{out}}}\rangle  = {\mu ^2}\langle {{{\hat n}_\textrm{in}}} \rangle  + {\nu ^2},
 \end{equation}
  \begin{equation}\label{eq15}
\langle {{{\hat n}_\textrm{out}}^2} \rangle  = {\mu ^4}\langle {{{\hat n}_\textrm{in}}^2}\rangle  + 3{\mu ^2}{\nu ^2}\langle {{{\hat n}_\textrm{in}}} \rangle  + {\mu ^2}{\nu ^2} + {\nu ^4},
 \end{equation}
  \begin{equation}\label{eq16}
 \begin{split}
\langle {{{\hat n}_\textrm{out}}^3}\rangle & = {\mu ^6}\langle {{{\hat n}_\textrm{in}}^3} \rangle  + 6{\mu ^4}{\nu ^2}\langle {{{\hat n}_\textrm{in}}^2} \rangle  + 4{\mu ^4}{\nu ^2}\langle {{{\hat n}_\textrm{in}}} \rangle \\&
 + 7{\mu ^2}{\nu ^4}\langle {{{\hat n}_\textrm{in}}} \rangle  + {\mu ^4}{\nu ^2} + 4{\mu ^2}{\nu ^4} + {\nu ^6}
\end{split},
 \end{equation}
   \begin{equation}\label{eq17}
 \begin{split}
\langle {{{\hat n}_\textrm{out}}^4} \rangle & = {\mu ^8}\langle {{{\hat n}_\textrm{in}}^4}\rangle  + 10{\mu ^6}{\nu ^2}\langle {{{\hat n}_\textrm{in}}^3} \rangle  + 10{\mu ^6}{\nu ^2}\langle {{{\hat n}_\textrm{in}}^2}\rangle \\& + 25{\mu ^4}{\nu ^4}\langle {{{\hat n}_\textrm{in}}^2} \rangle+ 11{\mu ^4}{\nu ^4} + {\mu ^6}{\nu ^2} + 11{\mu ^2}{\nu ^6} 
 \\&
 + {\nu ^8}  + ( {30{\mu ^4}{\nu ^4} + 5{\mu ^6}{\nu ^2} + 15{\mu ^2}{\nu ^6}} )\langle {{{\hat n}_\textrm{in}}} \rangle
\end{split},
 \end{equation}
where, $\mu  = \cosh g$, $\nu  = \sinh g$ and ${{\hat n}_{\textrm{in}}} = {{\hat a}_{\textrm{in}}}^\dag {{\hat a}_{\textrm{in}}}$.

\emph{Optical parametric amplified HBT interferometer}---From Eq.~(\ref{eq13}), we know that an OPA can boost the input photon number. Thus, we can apply OPAs to the HBT interferometer to obtain an amplified correlation signal we need. Our scheme is depicted in Fig.~\ref{fig3}. We place two OPAs at ${r_1}$ and ${r_2}$ and let the starlight pass through the OPAs first. The notations, ${\hat a_{j,\text{in}}}$ $({\hat b_{j,\text{in}}})$, ${\hat a_{j,\text{out}}}$ $({\hat b_{j,\text{out}}})$, corresponds to ${\hat a_\text{in}}$ $({\hat b_\text{in}})$, ${\hat a_\text{out}}$ $({\hat b_\text{out}})$ in Fig.~\ref{fig2}, where $j=1, 2$. The starlight is injected in the port of ${\hat a_{j,\text{in}}}$ and a vacuum state is injected in the port of ${\hat b_\text{j,in}}$. After the OPAs, we only detect the output of ${\hat a_\text{j,out}}$ and then combine the two output signals into the correlator. With the property that OPA can boost the input photon number, we obtain an amplified correlation function.

Next, we calculate the correlation function and the noise of the new system. Using \cref{eq14,eq15,eq16,eq17}, we can easily get the forms of $\langle \hat n_{k, \textrm{out}}^i \rangle $ and $\langle {\hat n_{k^{'}, \textrm{out}}^i} \rangle $ $(i=1, 2, 3, 4)$  after passing through the OPAs. Plugging these new values into Eq.~(\ref{eq4}) and Eq.~(\ref{eq6}), the correlation function reduces to
 
\begin{equation}\label{eq18}
C_\text{OPA}^{'} = 2( \mu^2 \bar{n} + \nu^2 )(\mu^2 \bar{m} + \nu^2 ) \cos ( {k{r_0}\phi } ),
\end{equation}
And, the noise to
\begin{equation}\label{eq19}
\begin{split}
& {(\Delta {C_\text{OPA}^{'}} )^2}= {\mu ^8}( \bar{n} + 13{\bar{n}^2} + 32{\bar{n}^3} + 20{\bar{n}^4} + \bar{m} + 13{\bar{m}^2})\\&
+ {\mu ^8}(32{\bar{m}^3} +20{\bar{m}^4}+ 28\bar{n}\bar{m} + 68{\bar{n}^2}\bar{m} + 68\bar{n}{\bar{m}^2}) \\&
+ {\mu ^8}( 42{\bar{n}^2}{\bar{m}^2} + 40\bar{n}{\bar{m}^3}+ 40{\bar{n}^3}\bar{m}  )+ {\mu ^2}{\nu ^6}( 93 + 73\bar{n} \\&
+ 77\bar{m} )+14{\nu ^8}+ {\mu ^6}{\nu ^2}( 2 + 51\bar{n} + 138{\bar{n}^2} + 88{\bar{n}^3} + 51\bar{m} )\\&+{\mu ^6}{\nu ^2}(  138{\bar{m}^2}+ 88{\bar{m}^3} + 216\bar{n}\bar{m} + 136{\bar{n}^2}\bar{m} + 136\bar{n}{\bar{m}^2} )\\&
 + {\mu ^4}{\nu ^4}( {46 + 199\bar{n} + 131{\bar{n}^2} + 195\bar{m} + 164\bar{n}\bar{m} + 131{\bar{m}^2}} ) .
\end{split}
\end{equation}

\begin{figure}[h]
\centering
\includegraphics[width=0.4\textwidth]{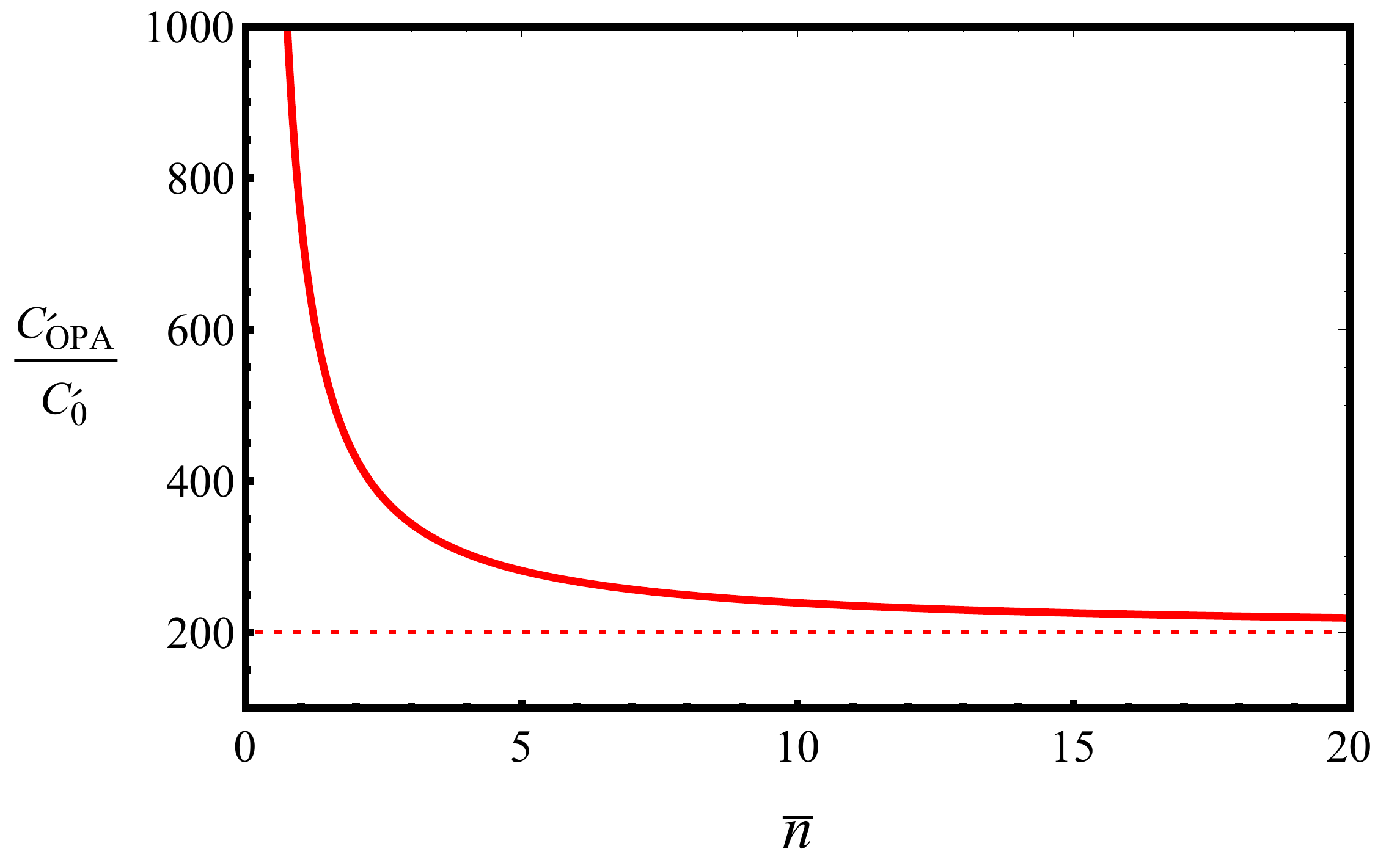}
\caption{\label{fig4} The ratio of the signals of the two interferometer schemes as a function of mean number of photons $\bar{n}$ in light rays $\vec {k}$ and $\vec k^{'}$, with $g = 2$. $C_\text{OPA}^{'}$ is the correlation function of the optical parametric amplified HBT interferometer and $C_0^{'}$ is the correlation function of the original HBT interferometer.}
\end{figure}

Comparing Eq.~(\ref{eq18}) with Eq.~(\ref{eq9}), we can see that the signal amplitude of the correlation function is increased by a factor of $(\bar{n}\cosh^2{g} + \sinh^2{g})( \bar{m}\cosh^2{g} + \sinh^2{g})/\bar{n}\bar{m}$. We plot the ratio of the signals of the two interferometer systems as shown in Fig.~\ref{fig4}. In this figure, we fix the parametrical strength $g=2$ and assume that both light sources have the same mean photon number. We see a factor of $( {{{\cosh }^2}g} )^2  \sim  200$ increase in the signal when $\bar{n}  \ge  10$.  

Once we have the signals and noises of both systems, that is of the original HBT interferometer and our proposed OPAs amplified HBT interferometer, we can compare the performance of these two systems. We use SNR as a indication of the system performance. As we discussed earlier, the signal is increased by the OPAs but what about SNR? 
We define the SNR as 
\begin{equation}\label{eq20}
\text{SNR} =  C^{'}/ {\Delta  C^{'}} .
\end{equation}

\begin{figure}[h]
\centering
\includegraphics[width=0.4\textwidth]{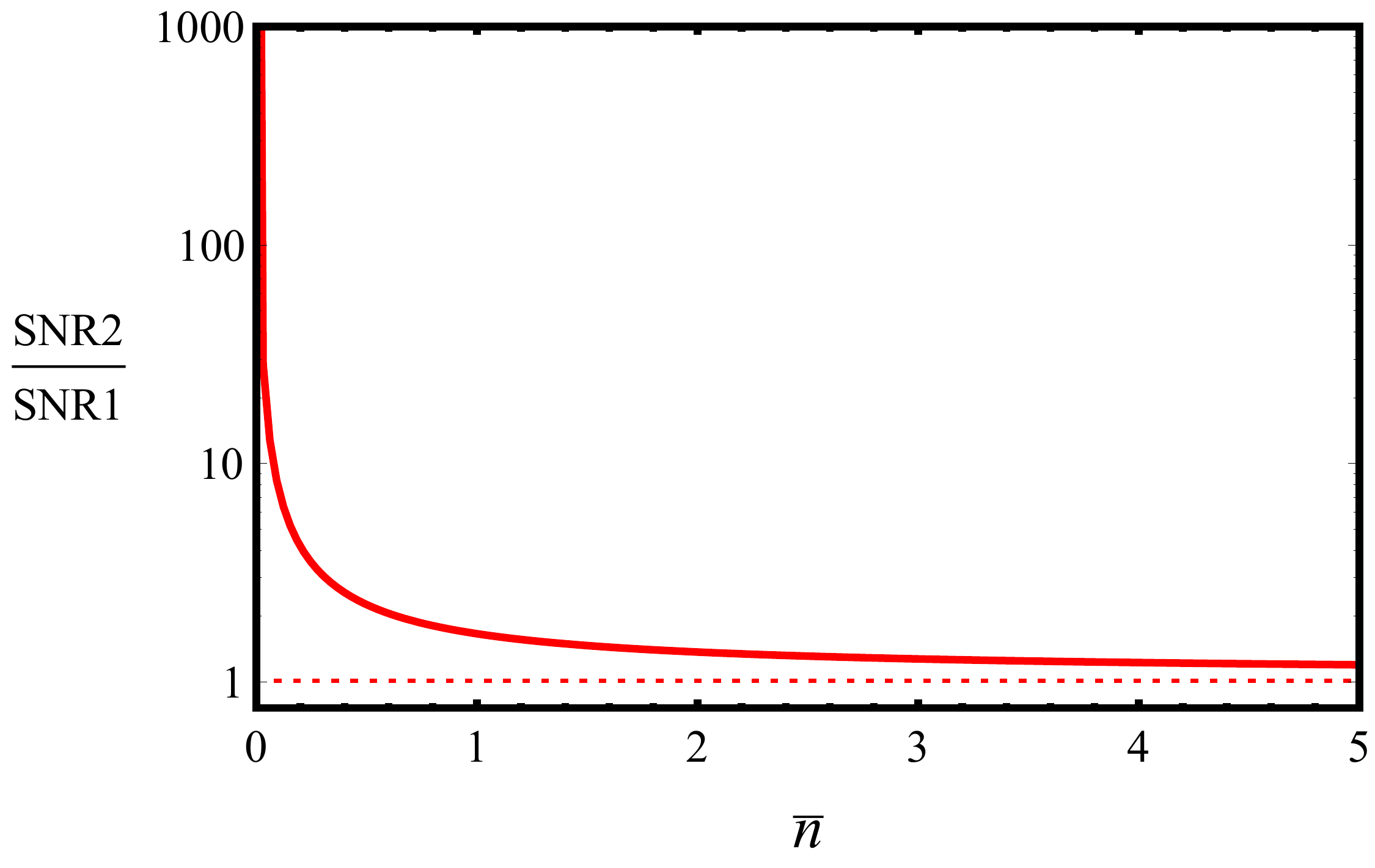}
\caption{\label{fig5} The ratio of the SNRs of the two interferometer schemes as a function of mean number of photons $\bar{n}$ in lights $\vec {k}$ and $\vec k^{'}$, with $g = 2$. SNR2 is the signal-to-noise ratio of the optical parametric amplified HBT interferometer and SNR1 is the signal-to-noise ratio of the original HBT interferometer.}
\end{figure}

In Fig.\ \ref{fig5}, we plot the ratio of the SNRs of the two systems as a function of the mean photon number. From the figure, we see that the SNR increases by about 66 percent with the use of the OPAs when $\bar{n} \approx 1$. Moreover, if $\bar{n}$ is much smaller, then the SNR increases by 400 percent. As the function of the curve is too complicated, we use curve fitting to get an approximated function which is 
\begin{equation}\label{eq21}
\frac{\text{SNR2}}{\text{SNR1}} = 1.082 + \frac{{0.584}}{\bar{n}}.
\end{equation}
From the above equation, one can get the value of SNR increased for an arbitrary small $\bar{n}$.

\emph{Conclusion}.---In conclusion, we have studied a new interferometric scheme that combines OPAs with HBT interferometers. In our scheme, instead of measuring the intensity of the starlight directly, we let the starlight go through the two OPAs first, amplifying the correlation signal by about a factor of ${\cosh ^4}g$. To be more specific, when the parametrical strength $g$ is 2, and the mean photon number of both star lights is 20, the signal gets 200 times larger than the original signal. Although the noise also increases in our new scheme, the SNR is increased by at least 8 percent compared to the original SNR. Even, when the starlight is weak, which means that the mean photon number is very small, the SNR increases by a factor of 400 percent. In astronomy, the intensity of light that reaches the detector from the majority of stars is very weak, thus, our scheme will be very helpful. For example, Vega which has $0.95 \times {10^{ - 4}}$ photons per unit optical bandwidth per unit area and unit time at 443nm as the flux is the second brightest star in the night sky \cite{Hanbury1974}. Using our interferometric scheme to measure Vega, the SNR could be increased by three orders of magnitude.

\emph{Acknowledgement}.---XPM acknowledges financial support from CSC, National Natural Science Foundation of China (Grant No. 61575180, 61701464, 11475160, 61640009) and the Natural Science Foundation of Shandong Province (Grants No. ZR2014AQ026 and No.ZR2014AM023). CY would like to acknowledge support from Economic Development Assistantship from Louisiana State University System Board of Regents. SA, HL and JPD would like to acknowledge support from the Air Force office of Scientific Research, the Army Research office, the Defense Advanced Research Projects Agency and the National Science Foundation.
 
\bibliographystyle{apsrev4-1}
\bibliography{HBT}

\end{document}